\documentclass[11pt]{article}
\usepackage{amssymb}

\newcommand{\bea}{\begin{eqnarray*}}
\newcommand{\eea}{\end{eqnarray*}}
\newcommand{\alfa}{\alpha}

\begin{document}

\title{The Birkhoff theorem for \\ unitary matrices of prime dimension}
\author{Alexis De Vos$^1$ and Stijn De Baerdemacker$^2$ \\[2mm]
           $^1$ Cmst, Imec v.z.w. \\
           vakgroep elektronica en informatiesystemen, \\ 
           Universiteit Gent, B - 9000 Gent, Belgium \\
           $^2$ Ghent Quantum Chemistry Group, \\
           vakgroep anorganische en fysische chemie, \\ 
           Universiteit Gent, B - 9000 Gent, Belgium} 

\maketitle

\begin{abstract}
The Birkhoff's theorem states that any doubly stochastic matrix lies
inside a convex polytope with the permutation matrices at the corners.
It can be proven that a similar theorem holds for unitary matrices with
equal line sums for prime dimensions.
\end{abstract}

\section{Introduction}

Doubly stochastic matrices are square matrices
with real entries, all belonging to the interval $(0,1)$,
such that all row sums and all column sums equal unity \cite{mirsky}.
Because the product of two doubly stochastic matrices
is again a doubly stochastic matrix, 
the doubly stochastic matrices form a \emph{semi}group.
They do not form a group 
because the inverse of a doubly stochastic matrix
is not necessarily a doubly stochastic matrix.
Because of their interpretation as probability distributions,
doubly stochastic matrices emerge in several sections of physics,
especially statistical physics.
Birkhoff's theorem~\cite{birkhoff} says that
any doubly stochastic matrix can be written as 
a weighted sum of permutation matrices,
such that all weights are real and belong to the interval $(0,1)$
and the sum of the weights equals unity.
So, every doubly stochastic matrix is contained in a convex set, 
spanned by the permutation matrices at the corners, thus dressing them with
a~geometric interpretation.
The higher-dimensional solid containing the matrix is called
Birkhoff's polytope~\cite{bengtsson}.

In the present paper, 
we aim to formulate an equivalent of Birkhoff's theorem
for unitary matrices.
The importance of unitary matrices equally follows from physics,
more in particular from quantum physics and quantum information.
In contrast to the $n \times n$ doubly stochastic matrices, 
the $n \times n$ unitary matrices form a genuine group,
called the unitary group and denoted U($n$).
Within this group figures a subgroup denoted XU($n$):
the group of $n \times n$ unitary matrices with 
all row sums and all column sums equal unity 
\cite{bremen} \cite{gent} \cite{acm}.
As such, XU($n$) acts as a `doubly stochastic' analogon within U($n$).
Whereas U($n$) is      an     $n^2$-dimensional Lie group,
       XU($n$) is only an $(n-1)^2$-dimensional Lie group,
isomorphic to U($n-1$).
Below, we will demonstrate Birkhoff-like properties for XU~matrices,
giving them a geometric interpretation.

\section{Three theorems}

\begin{quote}
{\bf Theorem 1}: If a U($n$) matrix can be decomposed
                 as a weighted sum $\sum_j m_jP_j$ of permutation matrices~$P_j$,
                 then it is, up to a global phase, member of the subgroup XU($n$).
\end{quote}
Indeed, let us assume that the matrix~$M$ can be written as $\sum_j m_jP_j$.
Each of the permutation matrices $P_j$ is a matrix with all line sums equal to~1.
Therefore the matrix $m_jP_j$ is a matrix with all line sums equal to~$m_j$, and
the matrix $\sum_j m_jP_j$ is a matrix with all line sums equal to $\sum_j m_j$.
If $M$ is member of U($n$) and has constant line sum, then this constant
can only be equal to a number of the form $e^{i\alfa}$, where $\alfa$ is an arbitrary real \cite{dyadic}.
Hence $M$ is of the form $e^{i\alfa}X$ with $X$ member of XU($n$). $\blacksquare$ \newline
Thus $M$ belongs to
the group of constant-line-sum unitary matrices $e^{i\alfa}X$, a group
isomorphic to the direct product U(1) $\times$ XU($n$) and thus
isomorphic to                    U(1) $\times$ U($n-1$).

Before introducing two more theorems, we present and prove two lemmas:

\begin{quote}
{\bf Lemma 1}: A circulant XU($n$) matrix can be written as a weighted sum of permutation matrices
               with the sum of the weights equal to~1.
\end{quote}
The proof is trivial, 
the decomposition consisting of the $n$~circulant $n \times n$ permutation matrices, 
each with a coefficient equal to one of the entries of the given XU~matrix. $\blacksquare$

\begin{quote}
{\bf Lemma 2}: If two matrices can both be written as a weighted sum of permutation matrices
               with the sum of the weights equal to~1, then also the product of the two matrices can.
\end{quote}
We consider two $n \times n$ matrices with the Birkhoff property, i.e.\ 
\[
a = \sum_u m_u^a \, P_u  \ \mbox{ and }\ 
b = \sum_v m_v^b \, P_v \ ,
\]
with
\[
\sum_u m_u^a = \sum_v m_v^b = 1 \ .
\]
Here, each $P_j$ denotes an $n \times n$ permutation matrix.
The product $c=ab$ is
\[
c = \sum_u \sum_v m_u^am_v^b \,  P_uP_v \ ,
\]
i.e.\ a matrix of the form $\sum_w m_w P_w$.
Because, moreover, $\sum_w m_w = \sum_u \sum_v m_u^am_v^b$ $= \sum_u m_u^a \ \sum_v m_v^b = 1$, 
we conclude that the product matrix~$c$ also has the Birkhoff property. $\blacksquare$

We now are in a position to present and prove the following theorem:
\begin{quote}
{\bf Theorem 2}: If a matrix belongs to XU($n$), 
then it can be written as a weighted sum of permutation matrices
with the sum of the weights equal to~1.
\end{quote}
The proof is by induction on~$n$:
we assume that the theorem is valid for $n=N$ and
consider an arbitrary matrix~$X$ from XU($N+1$).
It can be written as follows \cite{negator}:
\begin{equation}
X = F \left( \begin{array}{cc} 1 & \\ & U \end{array} \right) F^{-1} \ ,
\label{FUF}
\end{equation}
where $F$ is the $(N+1) \times (N+1)$ discrete Fourier transform
and $U$ is a matrix from U($N$). 
The matrix~$U$ can be written as follows \cite{gent} \cite{scaling} \cite{idel}:
\[
U = aZ_1xZ_2\ ,
\]
where $a$ is a member of U(1), i.e.\ a complex number with unit modulus,
where $x$ is a member of XU($N$), and
where both $Z_1$ and $Z_2$ are member of ZU($N$).
Here, ZU($n$) is the $(n-1)$-dimensional subgroup of U($n$),
consisting of all diagonal $n \times n$ unitary matrices 
with upper-left entry equal to unity
\cite{bremen} \cite{gent} \cite{acm}
and thus isomorphic to U(1)$^{n-1}$.
Because of our induction assumption, the matrix~$x$ can be written as
\[
x = \sum_j m_j p_j\ ,
\]
where all $p_j$ are $N \times N$ permutation matrices and $\sum_j m_j=1$.
We conclude that
\[
X = F \left( \begin{array}{cc} 1 & \\ & aZ_1 \sum_j m_j p_j Z_2 \end{array} \right) F^{-1} \ ,
\]
such that we have the matrix decomposition
\[
X = X_1YX_2\]
with
\bea
X_1 & = & F\, {\tiny \left( \begin{array}{cc} 1 & \\ & aZ_1         \end{array} \right)}\, F^{-1} \\[1mm] 
Y   & = & F\, {\tiny \left( \begin{array}{cc} 1 & \\ & \sum_jm_jp_j \end{array} \right)}\, F^{-1} \\[1mm]  
X_2 & = & F\, {\tiny \left( \begin{array}{cc} 1 & \\ & Z_2          \end{array} \right)}\, F^{-1} \ .
\eea
First, we note that both 
{\tiny $\left( \begin{array}{cc} 1 & \\ & aZ_1   \end{array} \right)$} and
{\tiny $\left( \begin{array}{cc} 1 & \\ &  Z_2   \end{array} \right)$} are
members of ZU($N+1$).
For any member $Z$ of ZU($N+1$) holds the property that $FZF^{-1}$ is
a circulant XU($N+1$) matrix. 
Thus, because of Lemma~1, both $X_1$ and $X_2$ 
can be written as a weighted sum of permutation matrices
(i.e.\ obey the theorem-to-be-proved).
Hence, by virtue of Lemma~2, to prove the theorem for~$X$, it suffices to prove it for~$Y$.
For this purpose, we note that, because of $\sum_jm_j=1$,
we have
\[
\hspace*{-8mm}
\left( \begin{array}{cc} 1 & \\ & \sum_jm_jp_j \end{array} \right) =
\left( \begin{array}{cc} \sum_jm_j & \\ & \sum_jm_jp_j \end{array} \right) =
\sum_j \left( \begin{array}{cc} m_j & \\ & m_jp_j \end{array} \right) =
\sum_j m_j \left( \begin{array}{cc} 1 & \\ & p_j \end{array} \right)  .
\]
We thus have
\[
Y = F\, \sum_j m_j    \left( \begin{array}{cc} 1 & \\ & p_j \end{array} \right)\, F^{-1} = 
        \sum_j m_j\ F \left( \begin{array}{cc} 1 & \\ & p_j \end{array} \right)\, F^{-1} \ .
\]
One can easily verify that any product of the form $F\, {\tiny \left( \begin{array}{cc} 1 & \\ & p_j \end{array} \right)}\, F^{-1}$
is a unitary matrix with upper-left entry equal to~1.
Because $F\, {\tiny \left( \begin{array}{cc} 1 & \\ & p_j \end{array} \right)}\, F^{-1}$ 
is of the form $F\, {\tiny \left( \begin{array}{cc} 1 & \\ & U \end{array} \right)}\, F^{-1}$,
this product is also an XU($N+1$) matrix \cite{negator}.
A~matrix with these two properties is necessarily of the form 
${\tiny \left( \begin{array}{cc} 1 & \\ & y_j \end{array} \right)}$
with $y_j$ a member of XU($N$).
Because of the induction hypothesis, we may put
\[
Y = \sum_j m_j\, \left( \begin{array}{cc} 1 & \\ & \sum_k m_k^yp_k  \end{array} \right) \ .
\]
Taking into account that $1=\sum_km_k^y$, we find
\[
Y = \sum_j m_j \sum_k m_k^y\ \left( \begin{array}{cc} 1 & \\ & p_k \end{array} \right) \ .
\]
Hence, $Y$ is of the Birkhoff form: a weighted sum of $(N+1) \times (N+1)$ permutation matrices 
with sum-of-weights equal to~1.
Hence, $X$ is. Thus the theorem holds for $n=N+1$.

All XU(2) matrices are of the form
\[
X = \frac{1}{2}\ \left( \begin{array}{cc} 1+e^{i\alfa} & 1-e^{i\alfa} \\ 1-e^{i\alfa} & 1+e^{i\alfa} \end{array} \right)\ .
\]
They can be written as a weighted sum of the two $2 \times 2$ permutation matrices:
\begin{equation}
X = \frac{1+e^{i\alfa}}{2}\ \left( \begin{array}{cc} 1 & 0 \\ 0 & 1 \end{array} \right) +
    \frac{1-e^{i\alfa}}{2}\ \left( \begin{array}{cc} 0 & 1 \\ 1 & 0 \end{array} \right) 
\label{n=2}
\end{equation}
and the sum of the two weights~$m_1$ and $m_2$ equals~1.

Because the theorem holds for $n=2$ and the theorem holds for $n=N+1$ as soon as it holds for $n=N$, 
the proof of Theorem~2 is complete. $\blacksquare$

Whereas the Birkhoff decomposition (\ref{n=2}) of an XU(2) matrix is unique,
the decomposition of an XU($n$) matrix with $n>2$ is not unique.
We now investigate whether, among the many possible decompositions $\sum_j m_jP_j$,
there is one or more that satisfies not only $\sum_j m_j=1$ but also $\sum_j |m_j|^2=1$.
This is a slightly stronger formulation of the $|m_j|<1$ constraints of 
the original Birkhoff theorem on doubly stochastic matrices
and again defines a convex polytope in which all XU($n$) lie.
We start with the case where $n$ is a prime:
\begin{quote}
{\bf Theorem 3}: If a matrix belongs to XU($n$) with prime~$n$, 
then it can be written as a weighted sum of permutation matrices
with the sum of the squared moduli of the weights equal to~1.
\end{quote}

Before proving Theorem~3, it is interesting to investigate 
some low-dimensional examples.
The theorem is trivial for $n=2$. Indeed, above, we have shown that
$m_1=(1+e^{i\alfa})/2$ and $m_2=(1-e^{i\alfa})/2$, such that
$|m_1|^2 +|m_2|^2=1$.

The theorem is also valid for $n=3$. In fact, there exist an infinity
of decompositions of $X$ as a weighted sum of the $n!=6$ permutation matrices,
all satisfying $\sum_{j=1}^6 |m_j|^2=1$. Indeed,
any member $X$ of XU(3) can be written as (\ref{FUF}),
with 
$F$ the $3 \times 3$ discrete Fourier transform and
$U$ a   $2 \times 2$ unitary matrix. Hence
\[
X = \frac{1}{\sqrt{3}}\ \left(\begin{array}{ccc} 1 & 1 & 1 \\ 1 & \omega   & \omega^2 \\ 1 & \omega^2 & \omega   \end{array}\right)\ 
                        \left(\begin{array}{ccc} 1 &   &   \\   & U_{11}   & U_{12}   \\   & U_{21}   & U_{22}   \end{array}\right)\ 
    \frac{1}{\sqrt{3}}\ \left(\begin{array}{ccc} 1 & 1 & 1 \\ 1 & \omega^2 & \omega   \\ 1 & \omega   & \omega^2 \end{array}\right) \ ,
\] 
where $\omega$ is the primitive 3~rd root of unity, i.e.\ 
$e^{i2\pi/3} = - \frac{1}{2} + i\, \frac{\sqrt{3}}{2}$.
The entries of $X$ therefore look like                           
\bea
X_{11} & = & (\ 1 +          U_{11} +          U_{12} +          U_{21} +          U_{22}\ )\, / \, 3 \\
X_{12} & = & (\ 1 + \omega^2 U_{11} + \omega   U_{12} + \omega^2 U_{21} + \omega   U_{22}\ )\, / \, 3 \\ 
X_{13} & = & (\ 1 + \omega   U_{11} + \omega^2 U_{12} + \omega   U_{21} + \omega^2 U_{22}\ )\, / \, 3 \\
X_{21} & = & (\ 1 + \omega   U_{11} + \omega   U_{12} + \omega^2 U_{21} + \omega^2 U_{22}\ )\, / \, 3 \\
...    &   & \\
X_{33} & = & (\ 1 +          U_{11} + \omega   U_{12} + \omega^2 U_{21} +          U_{22}\ )\, / \, 3 \ .
\eea
Each product $\omega^aU_{rs}$ ($\,\forall a = 0, 1, 2$ and $\forall r, s = 1, 2$)
appears exactly once in every row and exactly once in every column of~$X$.
Therefore it is staightforward to check that $X$ can be written as
%%%%% contributes to~$X$ by means of a term $\omega^aU_{rs}P_j$:
\bea
X & = & \frac{1}{3}\ [\ 
        (         U_{11} +          U_{22})\, P_1 + 
        (         U_{12} +          U_{21})\, P_2 + 
        (\omega   U_{12} + \omega^2 U_{21})\, P_3 + \\
  &   & (\omega^2 U_{11} + \omega   U_{22})\, P_4 + 
        (\omega   U_{11} + \omega^2 U_{22})\, P_5 +  
        (\omega^2 U_{12} + \omega   U_{21})\, P_6 \ ] + W_3 \ , 
\eea
 where
\[
P_1 = {\footnotesize \left( \begin{array}{ccc} 1 & 0 & 0 \\ 0 & 1 & 0 \\ 0 & 0 & 1 \end{array} \right)} , \ 
P_2 = {\footnotesize \left( \begin{array}{ccc} 1 & 0 & 0 \\ 0 & 0 & 1 \\ 0 & 1 & 0 \end{array} \right)} , \\[1mm]
P_3 = {\footnotesize \left( \begin{array}{ccc} 0 & 1 & 0 \\ 1 & 0 & 0 \\ 0 & 0 & 1 \end{array} \right)} , \ 
\]
\[
P_4 = {\footnotesize \left( \begin{array}{ccc} 0 & 1 & 0 \\ 0 & 0 & 1 \\ 1 & 0 & 0 \end{array} \right)} , \ 
P_5 = {\footnotesize \left( \begin{array}{ccc} 0 & 0 & 1 \\ 1 & 0 & 0 \\ 0 & 1 & 0 \end{array} \right)} , \ 
P_6 = {\footnotesize \left( \begin{array}{ccc} 0 & 0 & 1 \\ 0 & 1 & 0 \\ 1 & 0 & 0 \end{array} \right)} , \    
\]
and $W_3$ is the doubly stochastic matrix 
with all entries identical, i.e.\ equal to~$\frac{1}{3}$.
We call $W_3$ the $3 \times 3$ van der Waerden matrix \cite{waerden}.
It can be written both 
as a sum of the     circulant matrices and
as a sum of the anticirculant matrices:
\[
W_3 = \frac{1}{3}\ (P_1 + P_4 + P_5) = \frac{1}{3}\ (P_2 + P_3 + P_6) \ .
\]
Here, we apply the following decomposition:
\[
W_3 = \frac{1}{3}\ [\ p\, (P_1 + P_4 + P_5) + (1-p)\, (P_2 + P_3 + P_6)\ ] \ ,
\]
where $p$ is an arbitrary complex number.

Writing
\[
X = m_1P_1 + m_2P_2 +m_3P_3 + m_4P_4 + m_5P_5 + m_6P_6\ ,
\]
straightforward computations lead to $\sum_j m_j = 1$ and
\[
\sum_j m_j\overline{m_j} =  \frac{1}{3}\ [\ (U_{11}\overline{U_{11}} + U_{12}\overline{U_{12}}\, ) + 
                                            (U_{21}\overline{U_{21}} + U_{22}\overline{U_{22}}\, ) + 
                                            p\overline{p} + (1-p)(1-\overline{p}\, )\ ] \ .
\]
Taking into account that $U$ is a $2 \times 2$ unitary matrix leads to
\[
\sum_j m_j\overline{m_j} = 1 + \frac{1}{3}\ (2p\overline{p} - p - \overline{p}\, ) \ .
\]
For this sum to equal~1, it suffices that
\[
\left( p-\frac{1}{2}\right)\left( \overline{p}-\frac{1}{2}\right) = \frac{1}{4}\ ,
\]
i.e.\ that, in the complex plane, 
$p$~is located on the circle with center $\frac{1}{2}$ and radius $\frac{1}{2}$.
For the particular choice $p=1$, we obtain:
\bea
m_1 & = & (1 +          U_{11} +          U_{22}) / 3 \\
m_2 & = & (             U_{12} +          U_{21}) / 3 \\
m_3 & = & (    \omega   U_{12} + \omega^2 U_{21}) / 3 \\
m_4 & = & (1 + \omega^2 U_{11} + \omega   U_{22}) / 3 \\
m_5 & = & (1 + \omega   U_{11} + \omega^2 U_{22}) / 3 \\
m_6 & = & (    \omega^2 U_{12} + \omega   U_{21}) / 3 \ .
\eea

We are now in a position to prove Theorem~3 for an arbitrary prime.
We will suffice by demonstrating the existence of one appropriate decomposition.
Any member $X$ of XU($n$) can be written as (\ref{FUF}),
where $F$ is the $n \times n$ discrete Fourier transform
and $U$ is a matrix from U($n-1$). Hence, the matrix entries can be written
\[
X_{kl} = \frac{1}{n}\ [\ 1 + \sum_{r=1}^{n-1}\sum_{s=1}^{n-1} \omega^{(k-1)r-(l-1)s}\, U_{rs}\ ] \ ,
\]
where $\omega$ is the $n$~th root of unity.
Thus, given the numbers $r$ and $s$, each number $U_{rs}$
appears in the expression of every entry $X_{kl}$.
Therefore, we can write $X$ as a sum of $1+(n-1)^2$ matrices:
\[
X = W_n + \frac{1}{n}\ \sum_r \sum_s U_{rs}M_{rs} \ ,
\]
where $W_n$ is  the $n \times n$ van der Waerden matrix, 
i.e.\ the doubly stochastic matrix 
with all entries equal to $\frac{1}{n}$.
We call $M_{rs}$ the transfer factor of $U_{rs}$.
It is an $n \times n$ matrix with all entries equal to some $\omega^a$:
\[
\left( M_{rs}\right)_{kl} = \omega^{(k-1)r - (l-1)s}
\]
and with all line sums equal to~0.                                    

As $n$ is prime, a given number $\omega^a$ appears 
only once in every row and 
only once in every column of~$M_{rs}$.
Moreover $M_{rs}$ has the structure of a `supercirculant' matrix.
                A~square matrix~$A$ is called supercirculant if there exist two integers~$x$ and~$y$, 
                such that, for all $\{ k,l\}$, we have both 
                $A_{k+1,l+x} = A_{k,l}$ and $A_{k+y,l+1} = A_{k,l}$ (where sums are modulo~$n$).
                The numbers~$x$ and~$y$ (with $1 \le x \le n-1$ and $1 \le y \le n-1$) 
                are called the pitches. They are interdependent, as
                \[ xy = 1 \mbox{ mod } n \ . \]
                If $x=1$,   then $y=1$   and $A$ is simply called     circulant; 
                if $x=n-1$, then $y=n-1$ and $A$ is        called anticirculant.
The matrix $M_{rs}$ is supercirculant because
the difference $l(K+1) - l(K)$ in column number,
in which $\omega^a$ (for a~given~$a$) occurs for two consecutive rows~$K$ and $K+1$, 
is a constant (modulo~$n$) independent of~$K$. 
Indeed, applying $\omega^{(k-1)r-(l-1)s}$ equal to $\omega^a$
for both $k=K$ and $k=K+1$ yields
\[
(K-1)r - [\, l(K)-1\, ]\, s = Kr - [\, l(K+1)-1\, ]\, s \ \mbox{modulo}\  n
\]
and thus
\[
sl(K+1) - sl(K) = r \ \mbox{modulo}\  n \ ,
\]
such that    
$l(K+1)-l(K)$ is a constant, say~$x$. Analogously, for a given~$\omega^a$, 
$k(L+1)-k(L)$ is a constant, say~$y$.
We can summarize that the two pitches $x$ and $y$ follow from
\begin{eqnarray}
sx & = & r \mbox{ mod } n \nonumber \\
ry & = & s \mbox{ mod } n \ . 
\label{rs}
\end{eqnarray}
Because $n$ is prime, $x$ and $n$ are coprime and so are $y$ and~$n$.
Therefore, $\omega^a$ does not appear more than once in a column or row of~$M_{rs}$. 
As an example, for $n=5$, 
the eqns (\ref{rs}) yield the following functions $x(r,s)$ and $y(r,s)$:
\[
\begin{tabular}{c|rccc}
 $r\,\backslash\, s$ & $\ $1 & 2 & 3 & 4 \\[2mm]
\hline
      &   &   &   &   \\[-3mm]
    1 & 1 & 3 & 2 & 4 \\
    2 & 2 & 1 & 4 & 3 \\
    3 & 3 & 4 & 1 & 2 \\
    4 & 4 & 2 & 3 & 1 \end{tabular} \ \ \mbox{ and } \ \ 
\begin{tabular}{c|rccc}
 $r\,\backslash\, s$ & $\ $1 & 2 & 3 & 4 \\[2mm]
\hline
      &   &   &   &   \\[-3mm]
    1 & 1 & 2 & 3 & 4 \\
    2 & 3 & 1 & 4 & 2 \\
    3 & 2 & 4 & 1 & 3 \\
    4 & 4 & 3 & 2 & 1 \end{tabular} \ \ ,
\]
respectively.
From the table, one can read that the pitches of $M_{12}$ for $n=5$
are $x=3$ and $y=2$, respectively, leading to the explicit form
\[
M_{12} = 
       \left( \begin{array}{ccccc} 1        & \omega^3 & \omega   & \omega^4 & \omega^2 \\
                                   \omega   & \omega^4 & \omega^2 & 1        & \omega^3 \\
                                   \omega^2 & 1        & \omega^3 & \omega   & \omega^4 \\
                                   \omega^3 & \omega   & \omega^4 & \omega^2 & 1        \\
                                   \omega^4 & \omega^2 & 1        & \omega^3 & \omega   \end{array} \right) \ ,
\]
with $\omega$ equal to the 5~th root of unity, i.e.\ $e^{i2\pi/5} = (\sqrt{5}-1 + i\,\sqrt{10+2\sqrt{5}}\, )/4$.

We thus can conclude that any transfer matrix can be written as
\[
M_{rs} = \sum_{l=1}^n \omega^{-(l-1)s} C_{l,x(r,s)} \ .
\]
Here, $C_{l,x}$, with $1 \le l \le n$ and $1 \le x \le n-1$,
denotes the $n \times n$ supercirculant permutation 
matrix\footnote{Note that $C_{lx}$ is a permutation matrix
                if and only if $x$ and $n$ are coprime.} 
with a first-row unit entry in column~$l$ and a pitch equal to~$x$. 
In other words: we have $(C_{l,x})_{1,l} = (C_{l,x})_{2,l+x} =1$.  

We thus obtain the following decomposition:
\begin{eqnarray}
X & = & \frac{1}{n}\ \sum_{r=1}^{n-1} \sum_{s=1}^{n-1} U_{rs} \sum_{l=1}^n     \omega^{-(l-1)s} C_{l,x(r,s)} + W_n \nonumber \\
  & = & \frac{1}{n}\ \sum_{l=1}^n     \sum_{x=1}^{n-1} C_{lx} \sum_{s=1}^{n-1} \omega^{-(l-1)s} U_{r(s,x),s} + W_n \ ,
\label{PE}
\end{eqnarray}
where we thus sum over all $n(n-1)$ supercirculant permutation matrices $C_{lx}$. 
We note here that, because of eqns~(\ref{rs}), 
different values of~$s$ in (\ref{PE}) give rise to different values of $r(s,x)$.
%%% such that, for a given permutation matrix~$C_{lx}$, 
%%% all $U_{rs}$ have both indices different.

For $n \neq 2$, we may apply the following decomposition of the van der Waerden matrix: 
\[
W_n = \frac{1}{n}\ \sum_{j=1}^n D_j \ ,
\]
where the $n$~permutation matrices $D_j$ are chosen such that they have no 1s in common
and are not supercirculant, e.g.\  $D_j=Q^{j-1}D_1$, where
\[
Q = 
\left(\begin{array}{cccccccc} 0 & 1 & 0 & 0 & ...   & 0 & 0 & 0 \\
                              0 & 0 & 1 & 0 & ...   & 0 & 0 & 0 \\
                              0 & 0 & 0 & 1 & ...   & 0 & 0 & 0 \\
                              \vdots & & & & & & &              \\
                              0 & 0 & 0 & 0 & ...   & 0 & 1 & 0 \\
                              0 & 0 & 0 & 0 & ...   & 0 & 0 & 1 \\
                              1 & 0 & 0 & 0 & ...   & 0 & 0 & 0 \end{array} \right) 
\mbox{ and }
D_1 =
\left(\begin{array}{cccccccc} 1 & 0 & 0 & 0 & ...   & 0 & 0 & 0 \\
                              0 & 1 & 0 & 0 & ...   & 0 & 0 & 0 \\
                              0 & 0 & 1 & 0 & ...   & 0 & 0 & 0 \\
                              \vdots & & & & & & &              \\
                              0 & 0 & 0 & 0 &       & 1 & 0 & 0 \\
                              0 & 0 & 0 & 0 &       & 0 & 0 & 1 \\
                              0 & 0 & 0 & 0 &       & 0 & 1 & 0 \end{array} \right) \ ,
\]
the former being called the shift matrix.
Thanks to such choice, the matrix sets $\{ C_{lx}\}$ and $\{ D_j\}$ do not overlap and
the sum $\sum_j m_jP_j$ consists of two separate parts:
\[
X = \sum_{l=1}^n \sum_{x=1}^{n-1} m_{lx}C_{lx} + \frac{1}{n}\ \sum_{j=1}^n D_j \ .
\]
These parts have the following respective properties:
\begin{itemize}
\item The $n(n-1)$~weights $m_{lx}$ of the permutation matrices $C_{lx}$ equal a sum of $n-1$ products:
      \[
      m_{lx} = \frac{1}{n}\ \sum_{s=1}^{n-1} \ \omega^{-(l-1)s}\ U_{r(x,s),s}\ .
      \] 
      %% Each $U_{rs}$ appears once in $n$~different weights, each time with a different~$\omega^a$.
      %% Therefore $\sum_j m_j\overline{m_j}\, $ contains $\frac{1}{n^2}\,\sum_{a=0}^{n-1} \omega^a\omega^{-a} = \frac{1}{n}$
      %% times the contribution $U_{rs}\overline{U_{rs}}$.
      %% In each of these $n$~weights also appear other $U_{uv}$.
      %% Therefore $\sum_j m_j\overline{m_j}\, $ contains 
      %% $\frac{1}{n^2}\,\sum_a \sum_b \omega^a\omega^{-b}$
      %% times the contribution $U_{rs}\overline{U_{uv}}$.
      %% Because, after (\ref{PE}), we have $a=-(l-1)s$ and $b=-(l-1)v$, such that
      %% \[
      %% \sum_a \sum_b \omega^a\omega^{-b} = \sum_{l=1}^n \omega^{(l-1)(v-s)} \ .
      %% \]
      %% Because $\{ u,v\} \neq \{ r,s\}$ implies both $u \neq r$ and $v \neq s$, we have $v-s \neq 0$
      %% and therefore the sum is equal to~0.
      %% Therefore $\sum_j m_j\overline{m_j}\, $ contains no contributions $U_{rs}\overline{U_{uv}}$ with $\{ u,v\} \neq \{ r,s\}$.
      Therefore
      \bea \sum_{l\, x} m_{lx}\,\overline{m_{lx}} 
           & = & \frac{1}{n^2}\ \sum_{l\, x}\ 
           \sum_{s=1}^{n-1} \omega^{-(l-1)s}\ U_{r(x,s),s}
           \sum_{t=1}^{n-1} \omega^{+(l-1)t}\ \overline{U_{r(x,t),t}} \\
           & = & \frac{1}{n^2}\ \sum_{x=1}^{n-1} \sum_{s=1}^{n-1} \sum_{t=1}^{n-1}\ U_{r(x,s),s}\ \overline{U_{r(x,t),t}} 
                 \ \sum_{l=1}^n \omega^{(l-1)(t-s)} \ .
      \eea
      With $\sum_{l=1}^n \omega^{(l-1)(t-s)} = n\, \delta_{st}$, we obtain
      \[
      \sum_{l\, x} m_{lx}\,\overline{m_{lx}} = \frac{1}{n}\ \sum_{x=1}^{n-1} \sum_{s=1}^{n-1}\ U_{r(x,s),s}\ \overline{U_{r(x,s),s}} \ .
      \]
      As $U$ is an $(n-1) \times (n-1)$ unitary matrix, we have $\sum_{s=1}^{n-1}\ U_{rs}\ \overline{U_{rs}} = 1$.
      Hence
      \[
      \sum_{l\, x} m_{lx}\,\overline{m_{lx}} = \frac{1}{n}\ (n-1)\ .
      \] 
\item The $n$~weights $m_j$ of the permutation matrices $D_j$ equal $\frac{1}{n}$
      and therefore contribute to $\sum_j m_j\overline{m_j}\, $ with an amount $n$ times $|\frac{1}{n}|^2$ and thus $\frac{1}{n}$.
\end{itemize}
The two parts together thus give rise to 
\[
\frac{1}{n}\ (n-1)  + \frac{1}{n} = 1 \ . \ \ \blacksquare
\]

We note that the above construction does not work for the special case $n=3$ because, 
for $n=3$, the matrices $D_1$, $D_2$, and $D_3$ are, by coincidence, anticirculant.
Therefore, $D_1$, $D_2$, and $D_3$ coincide with $C_{12}$, $C_{22}$, and $C_{32}$, respectively, 
such that the above special-purpose construction for $n=3$ was necessary.
As a matter of fact, the proposed Birkhoff decomposition consists
of $n(n-1)$ matrices~$C_{lx}$ and $n$ matrices~$D_j$,
thus of a total of $n^2$~permutation matrices~$P_j$. Only for $n>3$,
the relation $n^2 \le n!$ is valid and there exist enough permutation matrices
to prove Theorem~3 in the generic way.

If $n$ is not prime, then not all transfer matrices~$M_{rs}$ are supercirculant
and the key property of the decomposition, proposed in the proof, is not fulfilled. 
If both $r$ and $s$ are coprime with~$n$, then $M_{rs}$ is supercirculant.
The other transfer matrices consist of identical blocks of size $b \times c$ with
\[
b = \frac{n}{\mbox{gcd}(n,r)} \ \mbox{ and } c = \frac{n}{\mbox{gcd}(n,s)} \ .
\]
E.g., for $n=4$, the $4 \times 4$ matrix~$M_{12}$ 
has two identical blocks of size $b \times c = 4 \times 2$:
\begin{equation}
M_{12} = 
\left( \begin{array}{llll} 1        & \omega^2 & 1        & \omega^2  \\
                           \omega   & \omega^3 & \omega   & \omega^3  \\
                           \omega^2 & 1        & \omega^2 & 1         \\
                           \omega^3 & \omega   & \omega^3 & \omega    \end{array} \right)\ ,
\label{M12}
\end{equation}
where $\omega$ here is the 4~th root of unity, i.e.\ $i$.

Whether Theorem~3 is also valid if $n$ is a composite number,
is left for further investigation. 
At least it is valid for the smallest non-prime, i.e.\ for $n=4$.
This can be verified by checking that the decomposition
\[
X = \sum_{j=1}^{24} m_jP_j
\]
where the weights~$m_j$ have the values as in the Appendix.

\section{A consequence}

As already mentioned in Section~2,
any $n \times n$ unitary matrix $U$ can be decomposed as
\[
U = e^{i\alfa}\ Z_1XZ_2 \ ,
\]
where $e^{i\alfa}$ is an overall phase factor, $X$ is an XU($n$) matrix, and both
$Z_1$ and $Z_2$ are ZU($n$) matrices.
Applying the fact that 
$X$ can be written as a weighted sum of         permutation matrices, we can conclude that 
$U$ can be written as a weighted sum of complex permutation matrices.
Here, we define a complex permutation matrix as a
unitary matrix having one and only one non-zero entry in every row and every column
\cite{barry} \cite{yu} \cite{chen}.

\section{Conclusion}

We have demonstrated that all matrices of the group $e^{i\alfa}$XU($n$) can be
written as a weighted sum of permutation matrices
and that, among the U($n$) matrices they are the only ones
that can be decomposed that way. 
The sum of the weights equals $e^{i\alfa}$.
We prove that the sum of the squared moduli of the weights can be made equal to unity
whenever $n$ is prime,
giving a convex geometric interpretation to the decomposition,
as in the standard Birkhoff theorem.
The case of non-prime~$n$ is left for further investigation.
%% The conjecture is proved for both $n$~prime and $n=4$.

\appendix

\section*{Appendix}

An arbitrary member $X$ of XU(4) may be written as $\sum_{j=1}^{24} m_jP_j$ with
\bea
m_1    & = & (U_{11} + U_{22} + U_{33}) / 4 \\
m_2    & = & 1 / 4 \\
m_3    & = & (U_{12} + U_{21} + U_{23} + U_{32} + i U_{12} - i U_{21} + i U_{23} - i U_{32}) / 8 \\
m_4    & = & (U_{21} + U_{23} + i U_{21} - i U_{23}) / 8 \\
m_5    & = & (U_{12} + U_{32} - i U_{12} + i U_{32}) / 8 \\
m_6    & = & (U_{13} + U_{31}) / 4 \\
m_7    & = & 1 / 4 \\
m_8    & = & (i U_{13} - i U_{31}) / 4 \\
m_9    & = & (- U_{12} - U_{32} + i U_{12} - i U_{32}) / 8 \\
m_{10} & = & (- U_{22} - i U_{11} + i U_{33}) / 4 \\
m_{11} & = & (- U_{12} + U_{21} + U_{23} - U_{32} - i U_{12} - i U_{21} + i U_{23} + i U_{32}) / 8 \\
m_{12} & = & (- U_{21} - U_{23} - i U_{21} + i U_{23}) / 8 \\
m_{13} & = & (- U_{21} - U_{23} - i U_{21} + i U_{23}) / 8 \\
m_{14} & = & (  U_{12} - U_{21} - U_{23} + U_{32} + i U_{12} + i U_{21} - i U_{23} - i U_{32}) / 8 \\
m_{15} & = & (- U_{13} - U_{31}) / 4 \\
m_{16} & = & (  U_{12} + U_{32} - i U_{12} + i U_{32}) / 8 \\
m_{17} & = & (- U_{11} + U_{22} - U_{33}) / 4 \\
m_{18} & = & 1 / 4 \\
m_{19} & = & (- U_{22} + i U_{11} - i U_{33}) / 4 \\
m_{20} & = & (- U_{12} - U_{32} + i U_{12} - i U_{32}) / 8 \\
m_{21} & = & (  U_{21} + U_{23} + i U_{21} - i U_{23}) / 8 \\
m_{22} & = & (- U_{12} - U_{21} - U_{23} - U_{32} - i U_{12} + i U_{21} - i U_{23} + i U_{32}) / 8 \\
m_{23} & = & 1 / 4 \\
m_{24} & = & (- i U_{13} + i U_{31}) / 4 \ , 
\eea
where the condition $\sum_j |m_j|^2=1$ is fulfilled.
Here, the $n!=24$ permutation matrices have been ordered `lexicographically' as follows:
\[
\footnotesize
P_1    = \left(\begin{array}{cccc} 1 & 0 & 0 & 0 \\
                                   0 & 1 & 0 & 0 \\
                                   0 & 0 & 1 & 0 \\
                                   0 & 0 & 0 & 1 \end{array}\right), \ 
P_2    = \left(\begin{array}{cccc} 1 & 0 & 0 & 0 \\
                                   0 & 1 & 0 & 0 \\
                                   0 & 0 & 0 & 1 \\
                                   0 & 0 & 1 & 0 \end{array}\right), \ 
P_3    = \left(\begin{array}{cccc} 1 & 0 & 0 & 0 \\
                                   0 & 0 & 1 & 0 \\
                                   0 & 1 & 0 & 0 \\
                                   0 & 0 & 0 & 1 \end{array}\right), \ ...\, , 
\normalsize
\]
\[
\footnotesize
P_{23} = \left(\begin{array}{cccc} 0 & 0 & 0 & 1 \\
                                   0 & 0 & 1 & 0 \\
                                   1 & 0 & 0 & 0 \\
                                   0 & 1 & 0 & 0 \end{array}\right), \ 
P_{24} = \left(\begin{array}{cccc} 0 & 0 & 0 & 1 \\
                                   0 & 0 & 1 & 0 \\
                                   0 & 1 & 0 & 0 \\
                                   1 & 0 & 0 & 0 \end{array}\right) \ .
\normalsize
\]
In this ordering, the supercirculant permutation matrices are
$C_{11}=P_1$,        $C_{13}=P_6$, 
$C_{21}=P_{10}$,     $C_{23}=P_8$, 
$C_{31}=P_{15}$, 
$C_{41}=P_{19}$, and $C_{43}=P_{24}$. 

\end{document}